\documentclass[12pt]{spieman}  
\usepackage{amsmath,amsfonts,amssymb}
\usepackage{color,soul}
\usepackage{graphicx}
\usepackage{setspace}
\usepackage{tocloft}

\title{HCF (HREXI Calibration Facility): Mapping out sub-pixel level responses from high resolution Cadmium Zinc Telluride (CZT) imaging X-ray detectors.}

\author[a, b, *]{Arkadip Basak}
\author[a]{Branden Allen}
\author[a]{Jaesub Hong}
\author[a]{Daniel P. Violette}
\author[a]{Jonathan Grindlay}
\affil[a]{Harvard University - Center for Astrophysics, 60 Garden Street, Cambridge, MA - 02138, USA.}
\affil[b]{Anton Pannekoek Institute, University of Amsterdam, Science Park - 904, 1098XH Amsterdam, NL.}

\cftpagenumbersoff{figure}
\cftpagenumbersoff{table} 
\begin{document} 
\maketitle

\begin{abstract}
The High Resolution Energetic X-Ray Imager (HREXI) CZT detector development program at Harvard is aimed at developing tiled arrays of finely pixelated CZT detectors for use in wide-field coded aperture 3-200 keV X-ray telescopes. A pixel size of $\simeq$ 600 $\mu m$ has already been achieved in the ProtoEXIST2 (P2) detector plane with CZT read out by the NuSTAR ASIC. This paves the way for even smaller 300 $\mu m$ pixels in the next generation HREXI detectors. This article describes a new HREXI calibration facility (HCF) which enables a high resolution sub-pixel level (100 $\mu m$) 2D scan of a 256 $cm^2$ tiled array of 2 $\times$  2 cm CZT detectors illuminated by a bright X-ray AmpTek Mini-X tube source at timescales of around a day. HCF is a significant improvement from the previous apparatus used for scanning these detectors which took $\simeq$ 3 weeks to complete a 1D scan of a similar detector plane. Moreover, HCF has the capability to scan a large tiled array of CZT detectors ($32cm \times 32cm$) at 100 $\mu m$ resolution in the 10 - 50 keV energy range which was not possible previously. This paper describes the design, construction, and implementation of HCF for the calibration of the P2 detector plane.   
\end{abstract}


{\noindent \footnotesize\textbf{*}Arkadip Basak,  \linkable{arkadip.basak@cfa.harvard.edu} }

\begin{spacing}{2}   

\section{Introduction}
\label{intro}

HREXI (High Resolution Energetic X-ray Imager) is a NASA supported program at Harvard to develop high resolution pixelated CZT detectors (pixel size currently 600$\mu m$, ultimately 300$\mu m$) for wide-field coded aperture imaging of X-ray emission from black holes (stellar to super-massive) and the astrophysics of extreme flaring sources and transients in 3-200 keV. The immediate goal of the program is to demonstrate the science and technology with an HREXI  SmallSat Pathfinder (HSP) mission that would enable a future SmallSat Constellation mission to continuously image the full sky\cite{josh1}. While, HREXI is still under development, its CZT detectors and electrical readout architecture are in large parts based on its predecessor ProtoEXIST2 (P2)\cite{Hong1, Hong3}.

\begin{figure}[htp]
\centering
\includegraphics[width = 0.7\textwidth]{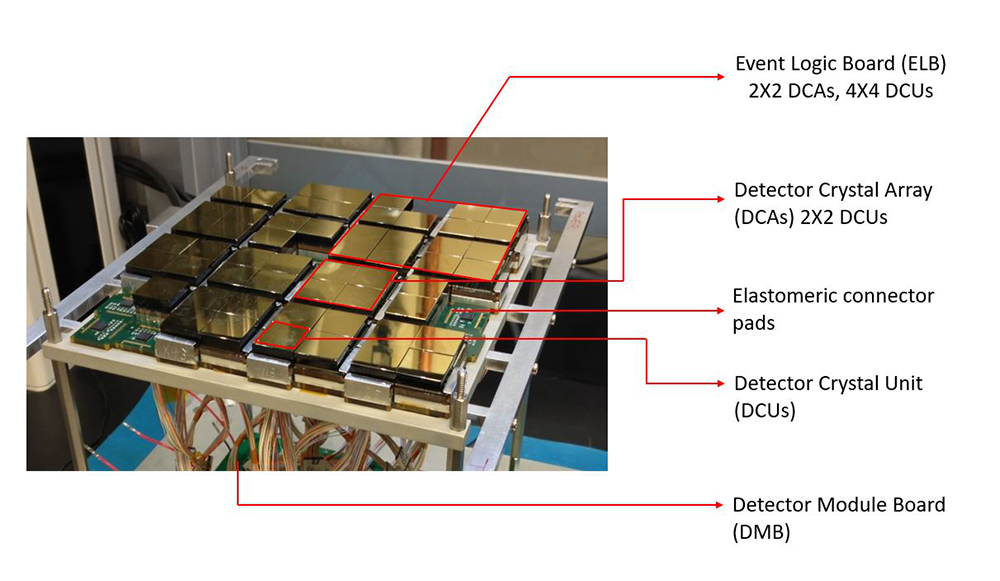}
\caption{Image of P2 detector plane showing DCUs, DCAs, ELBs and the DMB\cite{Hong2}.}
\label{intro2}
\end{figure}

P2 was a high altitude balloon-borne wide-field coded aperture hard X-Ray telescope with $\simeq$ 5 arcmin angular resolution. P2 consists of a tiled 8 $\times$ 8 array of pixelated detector crystal units (DCUs), where each DCU has a 19.9mm $\times$ 19.9mm $\times$ 5mm Redlen CZT crystal attached with a monolithic cathode and a 32 $\times$ 32 pixelated anode surrounded by a guard ring with adjustable bias bonded to an anode. Each anode pixel has a pitch of 604.8 $\mu m$ and an inter-pixel gap of $\simeq$ 50 $\mu m$. The anode pixels are directly bonded to an application-specific integrated circuit (ASIC) developed for NuSTAR (NuASIC)\cite{nustar}. The ASIC and its carrier board (ACB) maintain electrical contact through 87 individual wire-bonds. Groups of 2 $\times$ 2 DCUs are bundled together to form a Detector crystal Array (DCA). A 2 $\times$ 2 array of DCAs (4 $\times$ 4 array of DCUs) is mounted on an Event logic board (ELB)\cite{Hong2} which handles the low level ASIC interface logic. Thus, the entire P2 detector plane comprises of 56 DCUs (8 DCUs are currently not mounted in the P2 detector plane) and 4 ELBs. The 4 ELBs are in turn interfaced with the Detector Module Board (DMB) which serves as the primary point of communication between the detector plane and the flight computer. Figure \ref{intro2} shows the P2 detector plane along with its electrical architecture\cite{Hong2, Hong3}.

The 87 wirebonds between the ASIC and the ACB introduce a large $\simeq$ 14 pixel ($\simeq$ 8.4 mm) gap between CZTs as the wirebonds were fit into the ACB by extending it on one of the sides. The wirebonds are now being replaced by Through Silicon Vias (TSVs)\cite{josh1, dan1} for the next generation HREXI detectors. The use of TSVs will reduce this gap by more than a factor of 2. Moreover, TSVs will eventually enable the `3D ASIC' (TSV-NuASIC) to be flip-chip bonded to the readout electronics enabling lower cost and higher reliability production of large area CZT arrays\cite{josh1}.

\begin{figure}[htp]
\centering
\includegraphics[width = 0.7\textwidth]{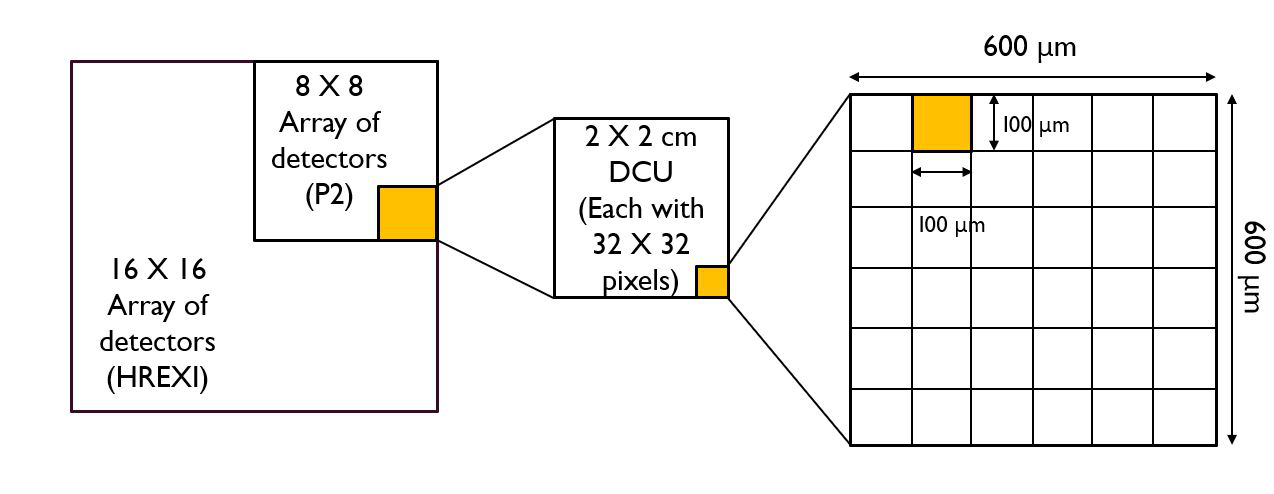}
\caption{Diagram showing the presence of $\simeq$ 56000 pixels and hence $>$ 2,000,000 sub-pixels in the P2 detector plane and the increase in that number by a factor of 16 for the future version of the HREXI detector plane with 300 $\mu m$ pixel pitch.}
\label{intro3}
\end{figure}

Allen et. al.\cite{subpix} have carried out a full P2 detector calibration revealing the existence of non-uniformities in CZT pixel maps which can be attributed primarily to the defects in the CZT crystals. Crystal defects lead to the migration of charge from pixel boundaries into adjacent pixels, effectively exhibiting under-sized or over-sized pixels in the P2 detector plane. This scan was carried out using a 10 mCi $^{241} Am$ source with a 1D fan beam having a collimated area of $\simeq$ 30 mm $\times$ 30 $\mu$m. However, low count rates from the $^{241} Am$ source leading to longer integration time ($\simeq 900$ seconds) and a large number of pixels ($\geq$ 56000) in the P2 detector plane meant a long exposure time ($\simeq$ 3 weeks) for completing such a scan. The exact number of pixels is less than 56 $\times$ 1024 because $\simeq$ 1000 pixels were identified as hot-pixels and disabled. The HREXI detector plane which will be used in the proposed 4$\pi$XIO array of SmallSats will have an even larger detector area with $16 \times 16$ array of DCUs and thus $\simeq$ 16 $\times$ 16 $\times$ 1024 pixels\cite{josh1}. For fine characterizations of detector non-uniformities at sub-pixel ($\simeq$ 100 $\mu m$) scales, it is essential to develop an apparatus that can scan such detector planes in much shorter timescales ($\simeq$ 1 - 3 days) and hence the development of HCF was undertaken. Figure \ref{intro3} illustrates the large number of sub-pixel scan points for both the P2 and the proposed future HREXI\cite{josh1} detector planes.

This article is organized in the following manner. Section \ref{sec1} provides an overview of the entire apparatus, while sections \ref{tube} to \ref{bottom_assem} describe each major component of the apparatus in detail. Section \ref{software} discusses the software development that was required for HCF commissioning, Section \ref{MiniX_perform} discusses the performance of the X-Ray source and Section \ref{initial_scan} shows initial scan results from the P2 detector plane. Section \ref{discussion} points to future directions and implications of the HCF for the development of HREXI.

\section{Experimental setup}
\label{sec1}

\begin{figure}[h]
\centering
\includegraphics[width = 0.9\textwidth]{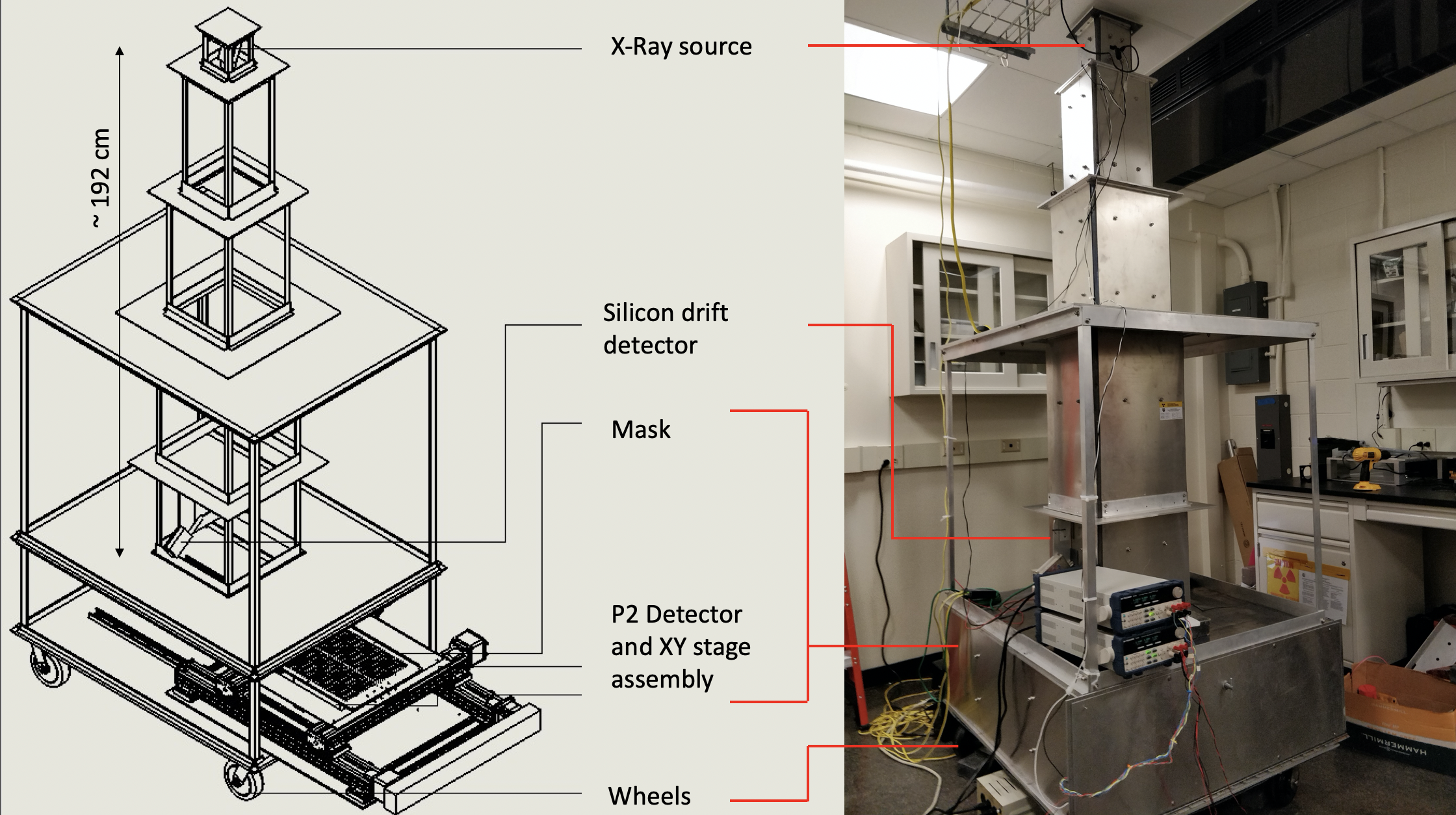}
\caption{HCF apparatus containing the X-Ray source, the Silicon Drift Detector, the mask, the XY stage assembly and the P2 detector plane.}
\label{exp1}
\end{figure}

The HCF apparatus is comprised of a bright X-Ray tube source (Section \ref{tube}) mounted on a tower like structure (Section \ref{alluminium}) which houses a silicon drift detector (Section \ref{SDD}) to monitor its X-ray flux. The bottom part of the assembly houses a Mask Assembly (Section \ref{Mask}), a precision XY stage assembly (Section \ref{XYstage}) and also the P2 detector box(Section \ref{bottom_assem}). The entire HCF apparatus is mounted on four heavy duty wheels at the bottom for ease of maneuverability. Figure \ref{exp1} (left) shows a schematic and figure \ref{exp1} (right) shows an actual image of HCF.

\subsection{X-Ray Tube source.}
\label{tube}

\begin{figure}[htp]
\centering
\includegraphics[width = 0.95\textwidth]{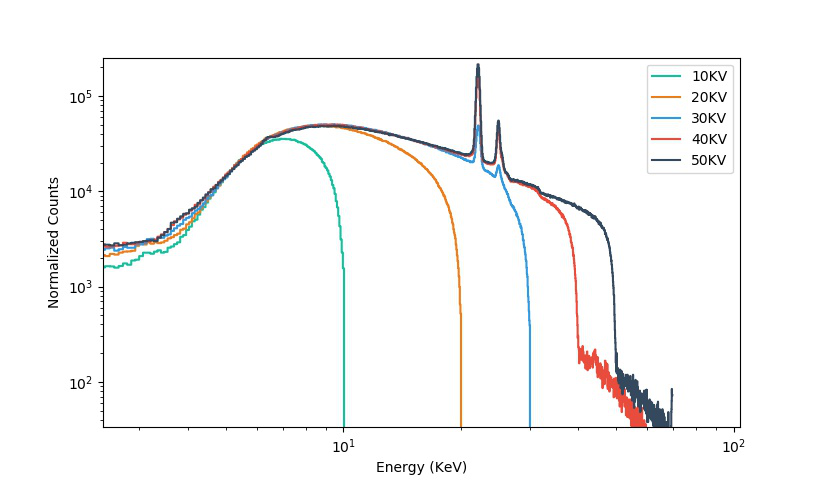}
\caption{Mini-X output spectrum (30 cms from source on axis) for different operating voltages\cite{MiniX}. An actual spectrum of Mini-X along with the fluorescent lines, as recorded by the silicon drift detector (X-123 SDD) is shown in figure \ref{MiniX_perform1}.}
\label{tube3}
\end{figure}

The `Mini-X' X-Ray tube system from AmpTek\cite{MiniX} is a self-contained miniature X-Ray tube system\cite{MiniX} that includes the X-Ray tube, the power supply, and a USB communication module which for control by an external computer. The Mini-X system using Silver (Ag) as a target from Amptek\cite{MiniX} has $K_{\alpha_1}$, $K_{\alpha_2}$ and $K_{\alpha_3}$ fluorescence lines at 21.71, 21.99 and 22.16 keV, $K_{\beta}$ lines at 24.8 keV, and a K edge at 25.52 keV plus a continuum spectrum from ~3 - 50 KeV depending on operating voltage. A thin Cu target can be inserted in the beam to produce 8.05 and 8.91 keV $K_{\alpha}$ and $K_{\beta}$ lines. These lines enable us to map and calibrate the detector response over these energies\cite{subpix}. Mini-X has an operating voltage between 10 and 50 kV and operating current between $5\mu A$ and $200\mu A$ with $4W$ maximum output power at $100 \%$ duty cycle. The beam from Mini-X has an opening angle of $120^{\circ}$ and flux of around $10^6$ counts $second^{-1} mm^{-2}$ at 50 KV operating voltage, $1\mu A$ current at a distance of 30 cm on axis\cite{MiniX}. Figure \ref{tube3} shows the normalized output spectrum of Mini-X at various operating voltages and $1\mu A$ operating current. The actual flux incident on P2 detector plane after attenuation through the Tungsten mask(see section \ref{Mask}) is estimated to be around $\simeq$ 1500 counts/sec. See section \ref{initial_scan} for more details.

\begin{figure}[htp]
\centering
\includegraphics[width = 0.9\textwidth]{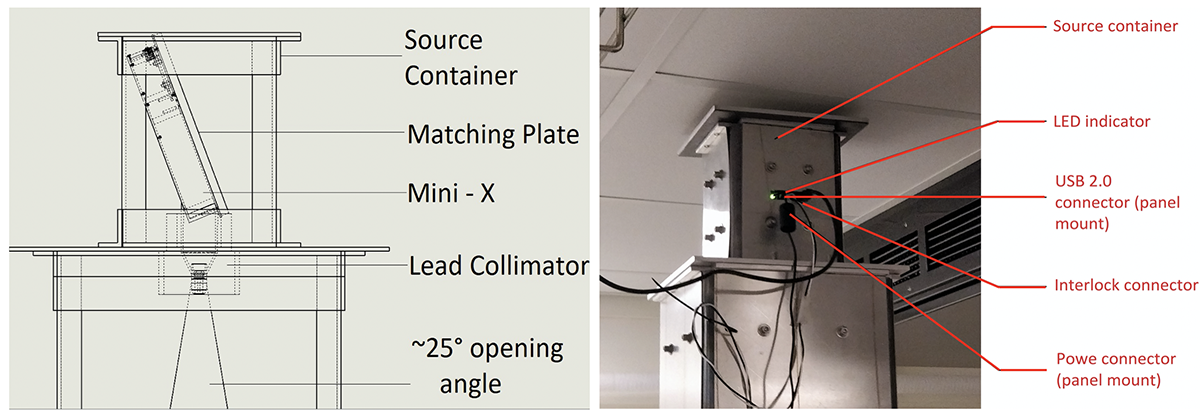}
\caption{Left: Schematic of Mini-X, the source container and the constrained opening angle for Mini-X output flux. Right: Mini-X source container housing the USB, power and the interlock connectors along with the LED indicator.}
\label{tube4}
\end{figure}

The Mini-X system is housed in a source container surrounded by lead sheets ($\simeq 1.5 mm$ thick) sandwiched between aluminum panels $1mm$ thick to prevent radiation leakage.  A lead collimator is placed at the bottom of the Mini-X to constrain the opening angle of the beam to $\simeq 25^{\circ}$ such that it illuminates the entire detector plane but reduces direct exposure on the tower side walls. This in turn reduces lead shielding requirements of the entire apparatus as seen in Section \ref{alluminium}. Figure \ref{tube4} (left) shows a schematic of the source, the attached matching plate and an opening angle of $25^{\circ}$ for Mini-X. One of the aluminum panels in the source holder also houses the power, USB, interlock (see section \ref{bottom_assem} for details) and an LED power indicator as shown in figure \ref{tube4}.  

\subsection{Aluminum frame and Lead shielding.}
\label{alluminium}

\begin{figure}[htp]
\centering
\includegraphics[width = 0.9\textwidth]{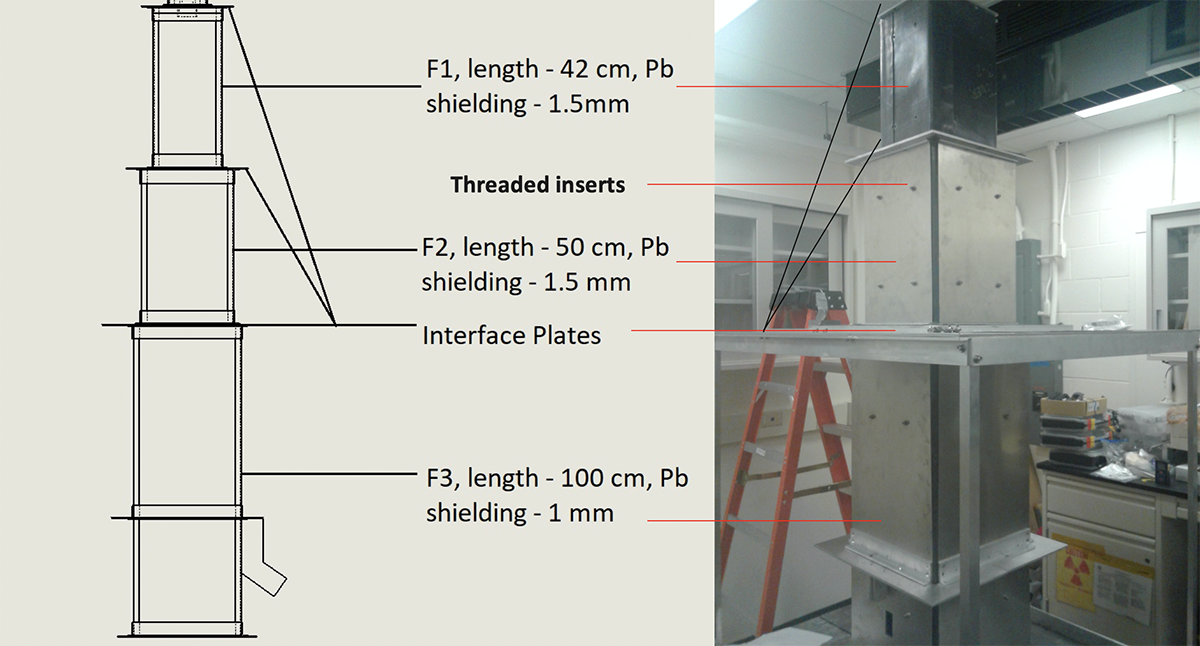}
\caption{Left: Schematic of F1, F2 and F3, the interface plates and the Lead shielding on each section of the frame. Right: Actual image of the frame assembly showing F1, F2, F3, interface plates and threaded inserts. The top layer of Aluminum panels have been removed from F1 to reveal the lead shielding underneath.}
\label{frame1}
\end{figure}

The source container is mounted to an interface plate which in turn is attached to an aluminum frame of length $\simeq 1.92 m$. The frame comprises of three parts (F1, F2 and F3) with interface plates attaching all three parts together (see Figure \ref{frame1}). A modular frame like this ensures that each part can be attached and removed separately during maintenance and troubleshooting. The modularity allows us to use different shielding thicknesses for each part of the frame thereby reducing the weight of the apparatus. 

 The normalized spectrum as shown in figure \ref{tube3} is used as a reference to calculate radiation dose rates from Mini-X. The Department of Environment Health and safety's (EH\&S)\cite{EHS} regulations limits require lead sheets of thicknesses between $\simeq$ 1 and 1.5 mm attached to the F1, F2 and F3 corresponding to a maximum dosage of 0.025 mrem per hour.

\subsection{SDD}
\label{SDD}

\begin{figure}[htp]
\centering
\includegraphics[width = 0.9\textwidth]{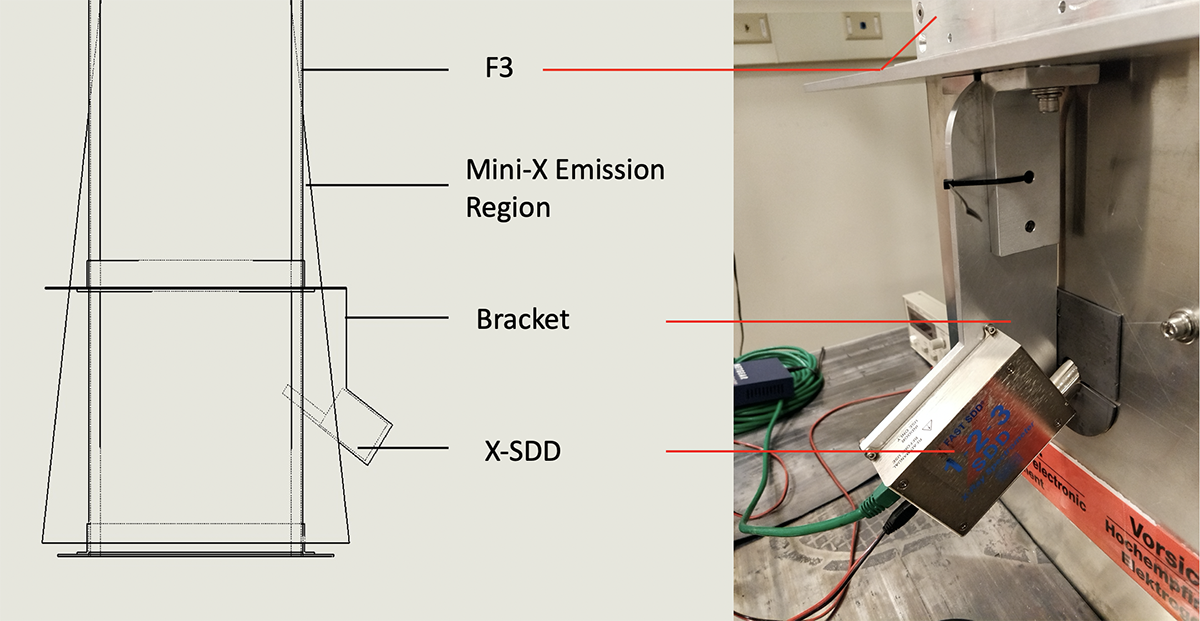}
\caption{Left: Schematic of X-123 SDD attached to F3 along with the constrained emission region of Mini-X. Right: Image of X-123 SDD. }
\label{sdd1}
\end{figure}

We have selected the X-123 SDD from Amptek\cite{sdd} as a Silicon Drift Detector X-Ray spectrometer to monitor the flux of Mini-X when the experiment is in progress. X-123 SDD is a single package containing XR-100SDD Silicon Drift diode\cite{sdd} and its charge sensitive pre-amplifier, the DP5 digital pulse processor (DPP) with pulse shaper, circuit amplifiers and the PC5 power supply (Refer to X-1233 SDD user-manual\cite{sdd} for technical details on DPP, pulse shaper and the power supply). The XR-100SDD can handle a maximum event rate of $\simeq 10^5$ counts per second\cite{sdd}. The PC5 supply requires a +5 VDC input and communications to the X-123 SDD from the computer were established through the Ethernet cable mountable on the back of the X-123 SDD (see Section \ref{software} for details). The attachment point of the X-123 SDD at F3 was chosen such that the X-123 SDD can peek directly at the Mini-X beam and the output flux from Mini-X is less than the X-123 SDD event rate limit\cite{sdd} at its attachment point as seen in figure \ref{sdd1}.

\subsection{Mask Assembly Design}
\label{Mask}
Tungsten Mask sheets with a repeating pattern of holes matched to the size of a single sub-pixel $ \simeq 100 \times 100 \mu m^2$ has been used to illuminate multiple sub-pixels of the detector plane simultaneously while shielding the other pixel regions from the Mini-X flux during calibration runs of the detector plane. The Mask (fabricated to our specs by the Tech Etch Co.\cite{tech}) is made out of layers of such 50 $\mu m$ thick laminated Tungsten sheets ($L_n$) with a hole pattern etched into them. Each hole is 100 $\mu m$ $\times$ 100 $\mu m$ as mentioned in section \ref{intro} such that it illuminates a $ \simeq 100 \times 100 \mu m^2$ area of a $600  \times 600 \mu m^2$ detector pixel. 

\begin{figure}[htp]
\centering
\includegraphics[width = 0.9\textwidth]{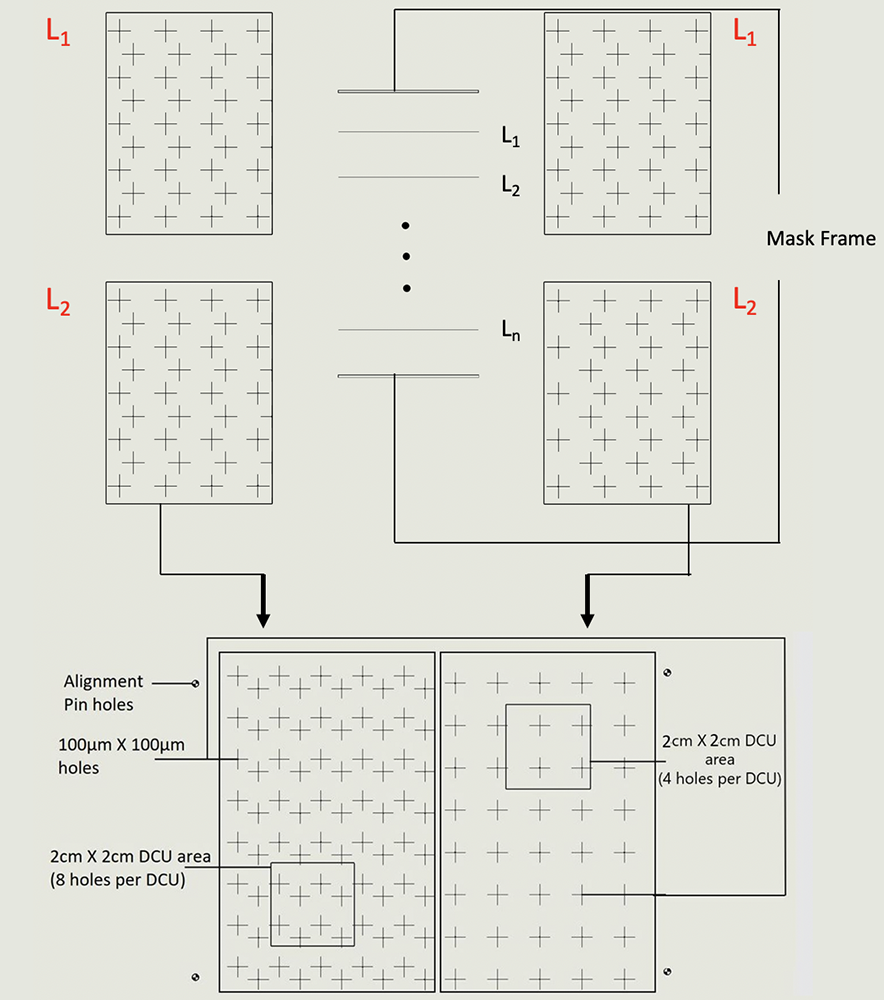}
\caption{Different possible orientations $L_1$ and $L_2$ leading to either 4 or 8 mask hole in a given 2 cm $\times$ 2 cm DCU area.}
\label{mask1}
\end{figure}

Each Tungsten sheet is $15.7 $ cm $\times$ $38.1$ cm in size. The identical hole pattern that has been etched onto them is such that the different orientations in which the layers $L_1, L_2, ..., L_n$ are stacked can give us a choice between either 4 or 8 holes (scan points) in  2 cm $\times$ 2 cm area, (the area of a single CZT crystal). Figure \ref{mask1} shows the different possible orientations in which the layers $L_1$ and $L_2$ can be arranged such that there are either 4 or 8 $100 \mu m \times 100 \mu m$ holes per DCU.

\begin{figure}[htp]
\centering
\includegraphics[width = 0.9\textwidth]{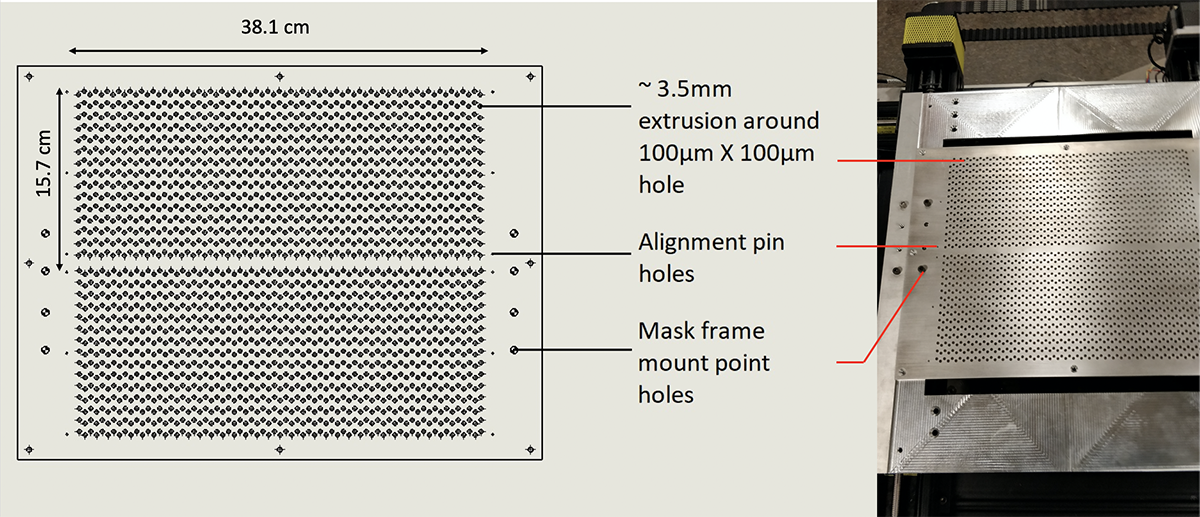}
\caption{Left: Schematic of the Mask frame displaying the larger $\simeq$ 3.5mm holes around each 100 $\mu m$ $\times$ 100 $\mu m$ hole, the alignment pin holes and the mount point holes. Right: Image of the assembled mask mounted on the XY-stage assembly.}
\label{mask4}
\end{figure}

The stacked layers of tungsten sheets $L_1, L_2, ..., L_n$ are then mounted on a frame (Mask frame) with the help of alignment pins. The frame comprises of two $1/8^"$ thick aluminum plates with the tungsten hole pattern matched into it but with significantly larger ($\simeq$ 3.5 mm in diameter) hole sizes. Thus, it provides rigid support around the thin 50 $\mu m$ tungsten sheets and the alignment pins ensure that the layers $L_1, L_2, ..., L_n$ are positioned accurately. Figure \ref{mask4} (left) shows a schematic of the frame while figure \ref{mask4} (right) shows an actual image of the Mask assembly.

%

\subsection{XY stage assembly}
\label{XYstage}

\begin{figure}[htp]
\centering
\includegraphics[width = 0.95\textwidth]{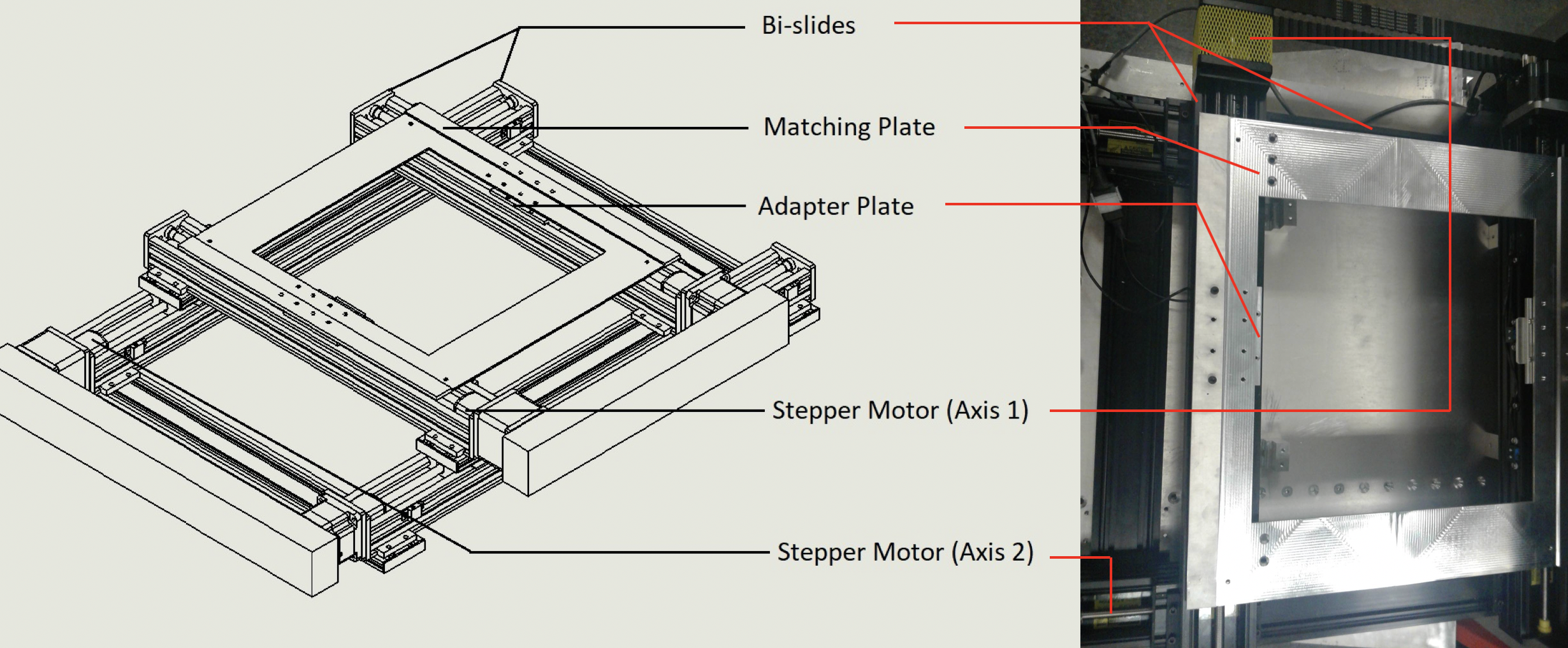}
\caption{Left: Schematic of the XY stage showing bi-slides, matching plate, adapter plate and stepper motors for both axes (X and Y). Right: Image of the XY stage. }
\label{stage1}
\end{figure}

HCF has been designed to map out spatial uniformities in CZT detector pixels at a sub-pixel ($100 \mu m$) scale for each 600 $\mu m$ $\times$ 600 $\mu m$ sized pixel in both X and Y directions. This requires a precision XY stage with X and Y stepping resolution $<<$ 100 $\mu m$. Moreover, it should be large enough to accommodate not only the current generation P2 detector plane but also the next generation HREXI detector plane (See section \ref{intro}). We use a parallel coupled bi-slide precision XY stage from Velmex\cite{xy} which has a resolution of $5 \mu m$ on both the X and Y axes along with a 33cm $\times$ 33cm opening in the center to accommodate both the P2 and the HREXI detector planes. Figure \ref{stage1} (left) shows a schematic of the stage while figure \ref{stage1} (right) shows an image of the assembled XY stage.

\begin{figure}[htp]
\centering
\includegraphics[width = 0.8\textwidth]{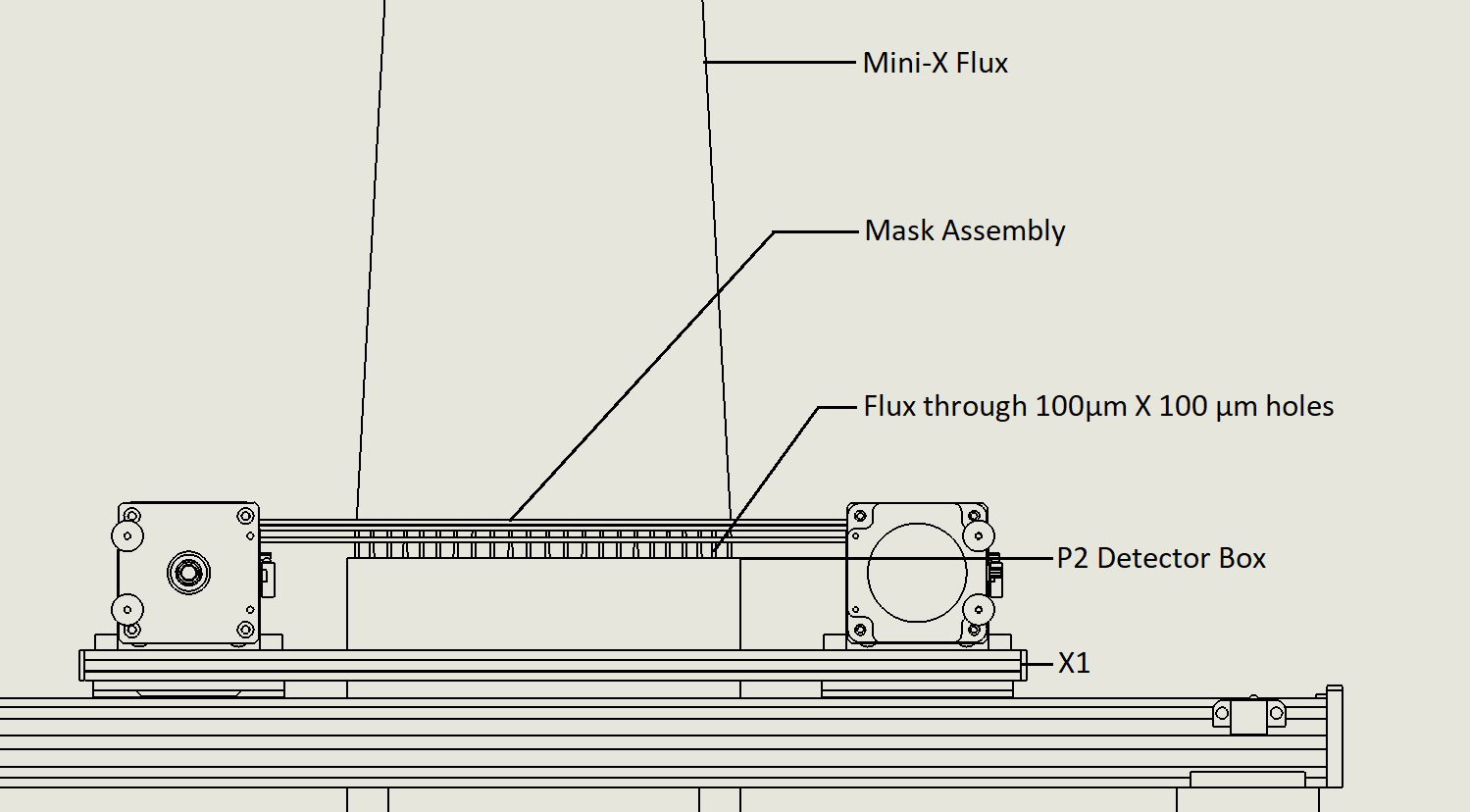}
\caption{Schematic of the stage, the assembled mask along with the simultaneous illumination of several 100 $\mu m$ $\times$ 100 $\mu m$ sub-pixels. }
\label{stage3}
\end{figure}

The bi-slides\cite{xy} of the stage have adapter plates mounted on top, which is interfaced to a matching plate on which the Mask Assembly can be mounted. Figure \ref{stage3} shows a cartoon diagram of the stage, the mask frame, the layers of tungsten sheets along with the illumination of several 100 $\mu m$ $\times$ 100 $\mu m$ sub-pixels by Mini-X simultaneously. There can be 4 or 8 mask holes per DCU depending on mask configuration (Section \ref{Mask}). For an 8 $\times$ 8 array of CZT detectors, the mask can illuminate 256 sub-pixels simultaneously. The simultaneous illumination of several sub-pixels also demonstrates how rapid sub-pixel level scanning can be achieved using HCF through parallel scanning of multiple detector regions.

\subsection{Bottom Assembly}
\label{bottom_assem}
\begin{figure}[htp]
\centering
\includegraphics[width = 0.9\textwidth]{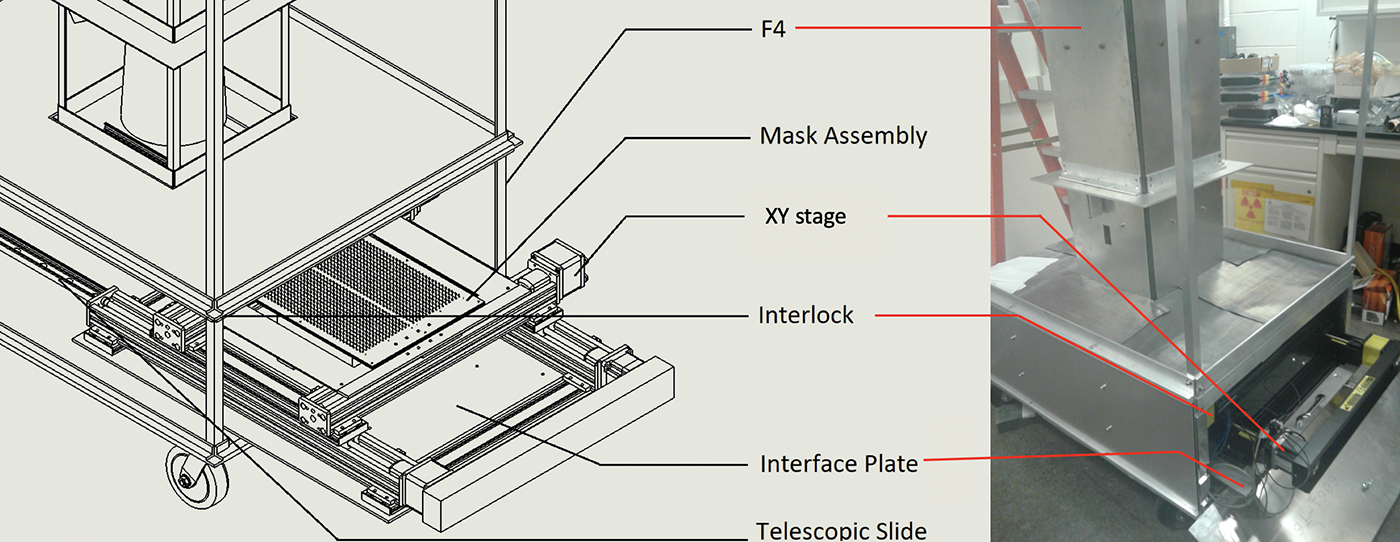}
\caption{Left: Schematic of the bottom assembly showing F4, mask assembly, XY stage, interlock and the telescopic slides. Right: Image of the bottom assembly.}
\label{bottom1}
\end{figure}

The XY stage, detector box along with the entire mask assembly is mounted to a plate which in turn is attached to a pair of heavy duty telescoping slides. The entire bottom assembly is housed in the bottom-most part of the apparatus (F4). F4 is designed in a similar manner to F1, F2 and F3 i.e. using aluminum L brackets, interface plates and surrounded by lead sheets sandwiched between aluminum panels on three sides. The fourth side houses a door that enables the entire XY stage assembly to slide out. The door and the slides ensure that the detector box and the Mask assembly can be accessed easily during maintenance or troubleshooting. F4 also houses an interlock which is directly interfaced with the power lines of Mini-X so that it can only be switched on once the door on F4 is shut to ensure safety. Figure \ref{bottom1} (left) shows a schematic of F4, the XY stage assembly, the telescoping slides and the position of the interlock, while figure \ref{bottom1} (right) shows an actual image.

\section{Software Development}
\label{software}
\subsection{X-123 SDD software development}
\label{X-SDD_software}
The X-123 SDD from Amptek\cite{sdd} comes with its own data acquisition and control software, DPPMCA\cite{sdd}. However, DPPMCA is not platform independent and can only be operated in Windows systems. Since, the communications and data acquisition from P2 detectors are handled by Linux-based systems\cite{Hong1, Hong2}, we have developed an in-house Linux-based python\cite{python} library which can communicate to the X-123 SDD through its Ethernet port. The library has built-in functions to retrieve spectral data, clear the spectrum buffer memory and receive status packets. The spectrum data from X-123 SDD comes in the form of 24 kbits data packets while status packets are 8 kbits each. The status packets contain data regarding the temperature, the detector accumulation and the dead times and hence is important to cross-correlate count rate values between the X-123 SDD and the P2 detectors. 

\subsection{XY stage and Scanning algorithm}
\label{scan}
P2 data reduction pipeline\cite{Hong1, Hong2, subpix} generates raw (uncompressed) event files which contain both event and housekeeping data of these detectors. We developed a Python library (Scan) that integrates the entire scanning process. It communicates with the XY stage via RS-232 serial communications port, the X-123 SDD via an Ethernet port and also manages the incoming P2 data-stream. 

\begin{figure}[htp]
\centering
\includegraphics[width = 0.95\textwidth]{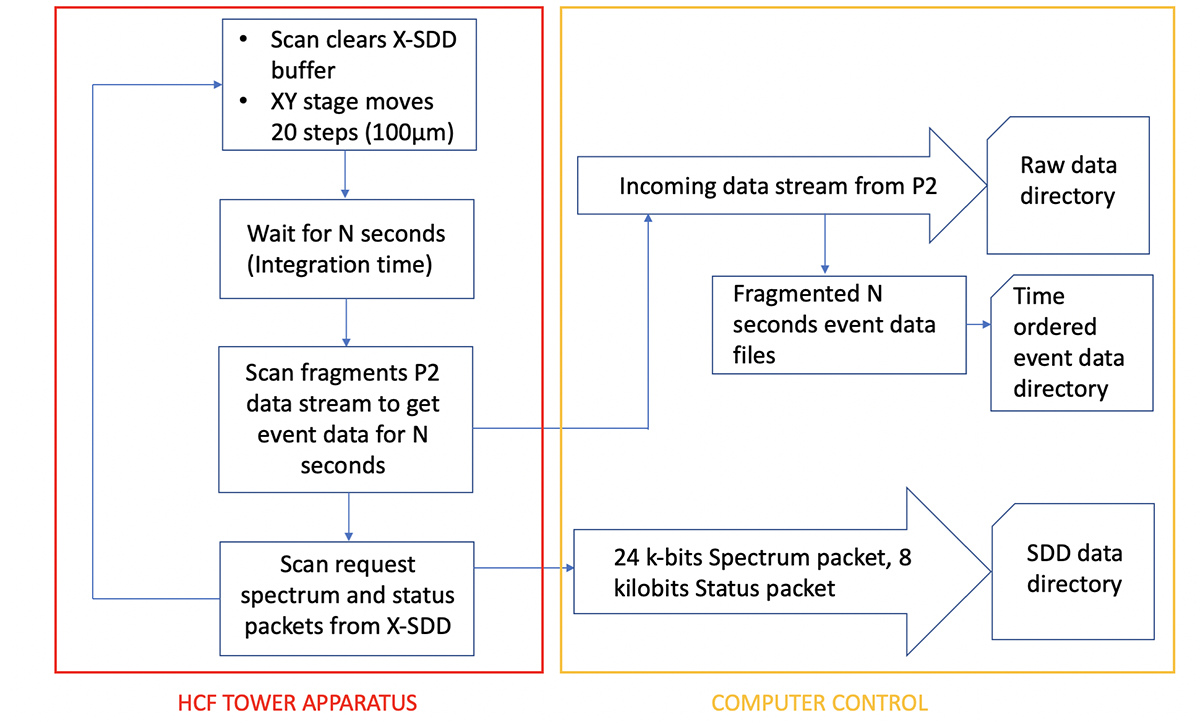}
\caption{Flowchart demonstrating the scanning algorithm and Scan library command control.}
\label{scan1}
\end{figure}

\begin{figure}[htp]
\centering
\includegraphics[width = 0.5\textwidth]{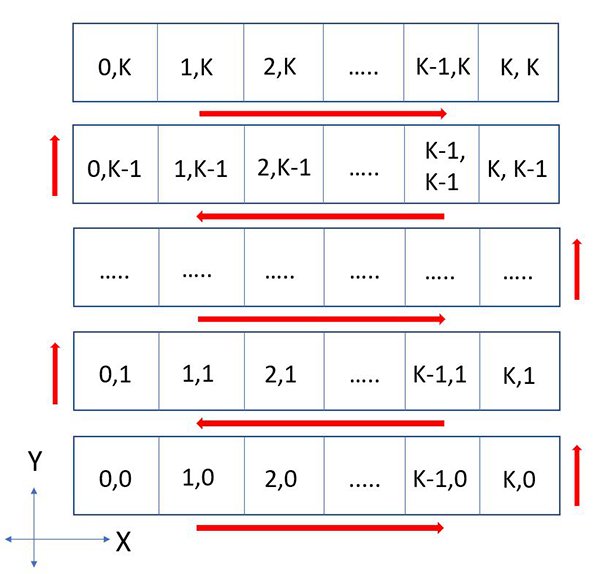}
\caption{Diagram of XY stage movement in X and Y coordinates. Each cell in the grid represents a 100 $\mu m$ $\times$ 100 $\mu m$ sub-pixel in the detector plane. The arrows represent the boustrophedon movement of the stage to scan the entire detector plane in K $\times$ K steps. The total number of scan steps K would depend on the number of $100 \mu m$ mask holes over each DCU.
}

\label{scan2}
\end{figure}

A flowchart for the scanning algorithm has been shown in figure \ref{scan1}. The incoming data-stream from P2 detectors is stored in a particular directory of the system (Raw). Scan sends commands to the XY stage to move 20 steps (100 $\mu m$) in the X direction. Simultaneously, it also clears the spectrum buffer memory of X-123 SDD and commands it to start collecting new spectrum data. After reaching the required integration time for each step, Scan fragments the P2 data-stream to get the event data for a given integration time (N seconds), receives the spectrum packet from X-123 SDD, requests a status packet from X-123 SDD and finally clears the X-123 SDD spectrum buffer. After moving for K times (K $\times$ 20 steps) in the +ve X direction (figure \ref{scan2}), Scan commands X1 to move 20 steps in the +ve Y direction followed by K $\times$ 20 steps in the -ve X direction and so on thereby covering the entire detector plane in a boustrophedon pattern. Figure \ref{scan2} shows a schematic of the XY stage movement to scan the entire detector plane.

\section{Performance of the Tube Source (Mini-X)}
\label{MiniX_perform}

\begin{figure}[htp]
\centering
\includegraphics[width = 0.95\textwidth]{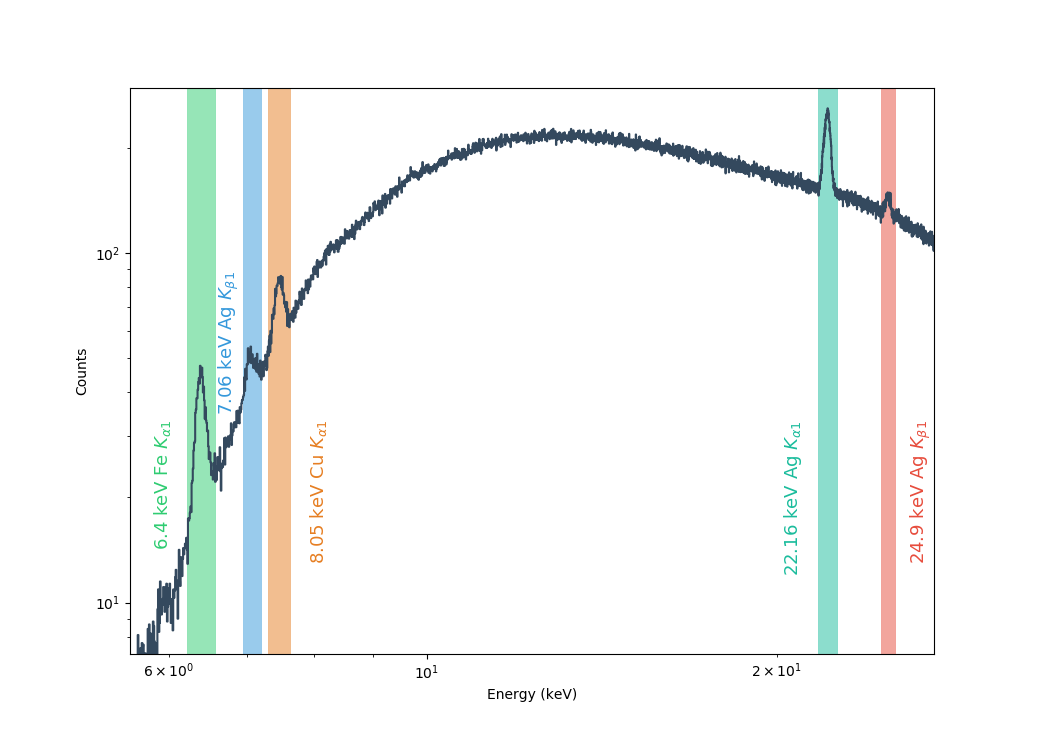}`
\caption{Output spectrum from the tube at 30 KV/130 $\mu A$ operating conditions as measured by the X-123 SDD. Prominent Ag fluorescence lines are visible above 22 keV together with fluorescent lines at lower energies which will enable calibration of the sub-pixel response HREXI (See section \ref{intro}). The presence of a Fe $K_{\alpha 1}$ line at 6.4 keV can be attributed to the steel frame of Mini-X\cite{MiniX}.}
\label{MiniX_perform1}
\end{figure}

Mini-X was operated at 30 KV/130 $\mu A$ and the output spectrum was observed using the X-123 SDD. The X-123 SDD spectrum was previously calibrated using an $^{241}Am$ source and also using the 22.16 Kev and 24.94 Kev Ag $K_{\alpha1}$ and $K_{\beta1}$ lines from Mini-X. The output spectrum with the major peaks is shown is figure \ref{MiniX_perform1}.

\begin{figure}[htp]
\centering
\includegraphics[width = 0.95\textwidth]{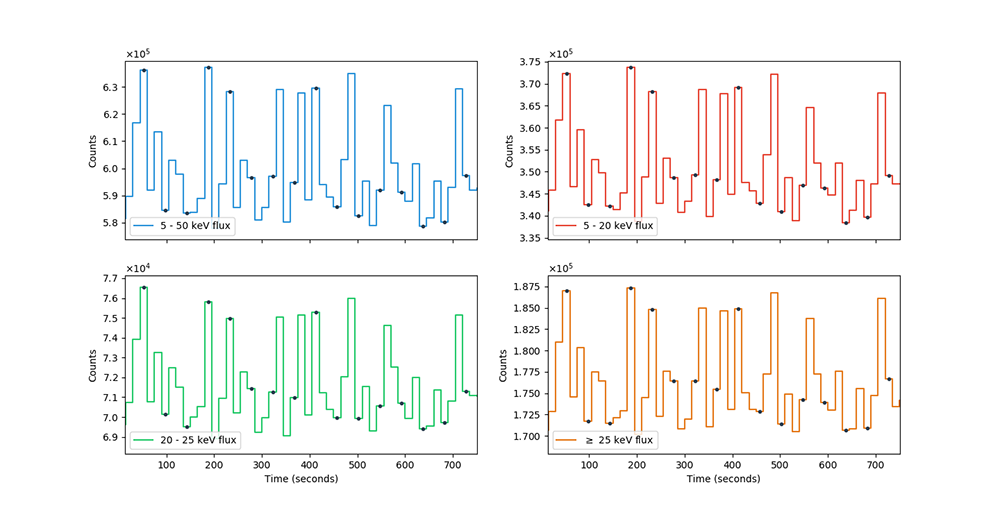}
\caption {Mini-X flux as measured from X-123 SDD. Top left: Evolution of 5 - 50 keV flux from Mini-X at 30 KV/130 $\mu A$ operating conditions. Top right: Evolution of 5 - 20 keV flux, Bottom left: Evolution of 20 - 25 keV flux, Bottom right: Evolution of $\geq$ 25 keV flux at similar operating conditions. The black dots represent the standard $1\sigma$ errors on count rates on some representative points.}
\label{MiniX_perform2}
\end{figure}

\begin{figure}[htp]
\centering
\includegraphics[width = 0.99\textwidth]{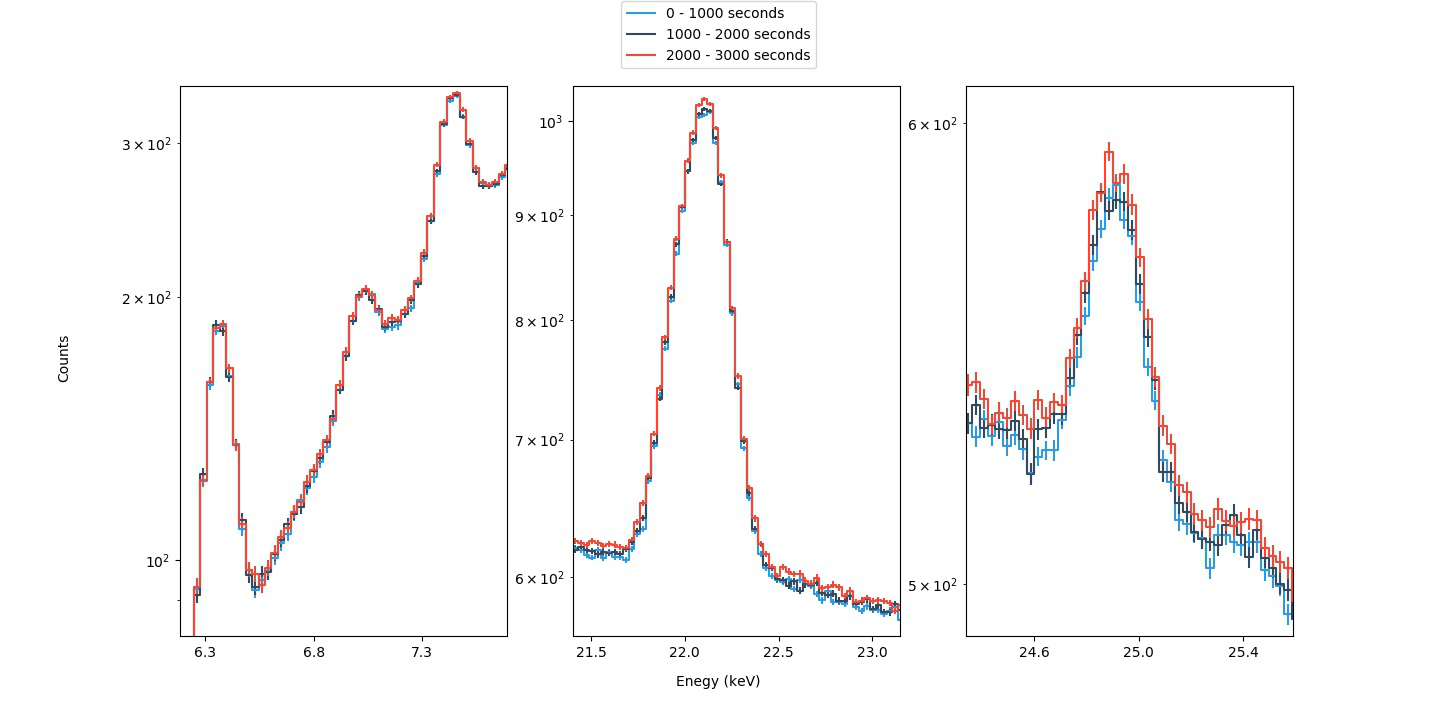}
\caption{Variation in the flux and position of the Fe, Cu and Ag peaks over longer timescales ($\gg$ Integration time for each scan step)  at 30 KV/130 $\mu A$ operating conditions. The error bars represent standard 1$\sigma$ errors.}
\label{MiniX_perform5}
\end{figure}

The variability of the output flux from Mini-X can be observed in figure \ref{MiniX_perform2} (top left). In order to confirm that these fluctuations in the total count rate are seen over the entire 5 - 50 keV energy range, the flux from Mini-X at different energy ranges are observed. Figure \ref{MiniX_perform2} confirms that these fluctuations are observed in all energies - the continuum at lower energies 5 - 20 keV (top right), around the Ag peaks 20 - 25 keV (bottom left) and the continuum at energies $\geq$ 25 keV. In addition to this, for each run of $\simeq$ 1000 seconds an averaged spectrum was obtained, calibrated, and the positions of the Fe, Cu, and Ag lines were investigated. The mean fluxes, as well as the positions of the peaks, remain constant over all the 1000 second intervals as shown in \ref{MiniX_perform5}

Figure \ref{MiniX_perform2} (Top left) shows clearly that the output flux from Mini-X is not constant and there can be variations in count rates over long time scales even though the mean flux and the position of the peaks (figure \ref{MiniX_perform5}) remain constant over 1000 second intervals. Since, we are interested in mapping out detector response at a sub-pixel level and a single run of the experiment can last up to $\simeq$ 3 days, it is essential to have the X-123 SDD to provide an absolute flux calibration for each scan step. The independent measurement of total output flux between the X-123 SDD and the P2 detectors would provide a one-to-one normalization factor for each step in the scan as seen in section \ref{discussion} and in section 2 of Ref. ~\citenum{basak1}. Once normalized for a constant Mini-X flux\footnote{See figures in Section 2 of Ref ~\citenum{basak1}.}, the residual variability in counts across sub-pixels can be attributed to P2 detector response\cite{basak1} and not changes in Mini-X flux.

\section{HCF commissioning and initial scan data.}
\label{initial_scan}

Once the fabrication of the HCF hardware was completed and all the software tools were developed, the P2 detector plane was installed in F4 beneath the mask. The mask was assembled using the mask frame along with $L_1, L_2$ and $L_3$ i.e. using three of the Tungsten sheets in a manner shown in figure \ref{mask1} (right) which allows four $100 \mu m \times 100 \mu m $ apertures over a single DCU. The operating conditions for Mini-X were 30 KV, 130 $\mu A$. The number of Tungsten sheets in the mask, its orientation and the operating conditions of Mini-X were chosen to limit $1 \sigma$ errors on the counts to $\simeq$ 3 \% for each scan step in a $\simeq$ 15 seconds integration time corresponding to $\simeq$ 1000 counts per sub-pixel. Operating Mini-X at a higher current, changing the orientation of mask layers to the one described in figure \ref{mask1} (left) are ways to reduce the integration time for each scan step and hence ensure a quicker scan. However, the P2 detector readout electronics cannot handle event rates beyond $\simeq 15000$ counts/second primarily due to back-end electronics, constraining the maximum operating current. The addition of tungsten sheets to the mask assembly increases the attenuation of Mini-X flux in the parts of the detector shielded by it, thereby reducing the event rate and hence allowing for higher Mini-X operating current (and a quicker scan). But, the problem with this method arises from the fact that since the hole pattern was etched into these Tungsten sheets, not all the holes are of the same size. There are variations in hole sizes ranging from $\simeq 80 \mu m$ to $\simeq 120 \mu m$ and adding layers to the mask assembly only compounds this issue. These variations were measured using a microscope after placing the Tungsten sheets on a light table.

\begin{figure}[htp]
\centering
\includegraphics[width = 0.95\textwidth]{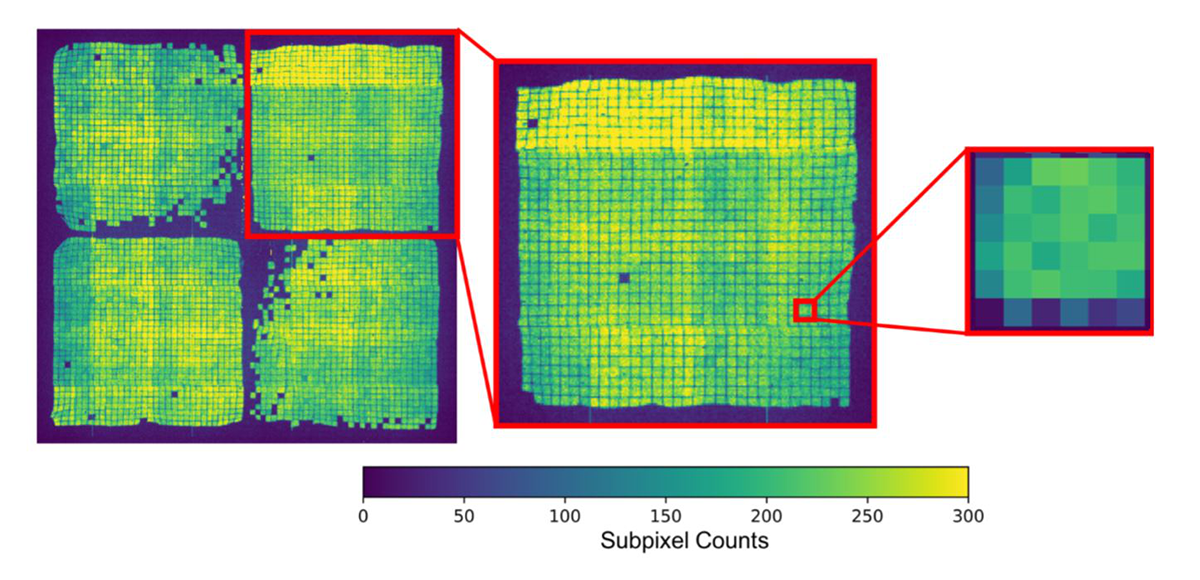}
\caption{From left to right: A DCA map comprising of 2 $\times$ 2 DCUs, a DCU containing 32 $\times$ 32 $\times$ 600 $\mu m$  $\times$ 600 $\mu m$ pixels and each pixel containing 6 $\times$ 6 $\times$ 100 $\mu m$  $\times$ 100 $\mu m$ sub-pixels. This DCA map was reconstructed using the modified version of P2 data analysis pipeline\cite{subpix, basak1}.}
\label{pixel_map}
\end{figure}

The P2 detector plane settings including low energy threshold settings for each DCU, hot-pixel disable maps and the detector temperature ($\simeq 25^{\circ}C$) were optimized for 2D scanning and a complete 2D scan into $100 \mu m$ sub-pixels of the entire detector plane was carried out lasting for $\simeq$ 60 hours. A full 2D map of each pixel was reconstructed using the P2 data analysis pipeline\cite{subpix, Hong2, Hong1}. This pipeline maps out the response of each individual sub-pixel a function of the XY stage movement. Figure \ref{pixel_map} shows a composite DCA map reconstructed using this pipeline and then zoomed in to show the reconstructed map of one example pixel. The variation in hole-sizes in the mask becomes much more apparent through light and dark patches in the DCU image showing the portions of DCUs scanned by different hole sizes and hence exhibiting unequal fluxes. The other patterns observed in the DCA maps are primarily due variations in Mini-X flux and their recurrence in the maps demonstrates the simultaneous scanning of multiple sub-pixels as the same flux pattern is observed in multiple sections scanned by different holes.

\section{Discussion and Future work}
\label{discussion}
Figure \ref{pixel_map} (left) shows a full 2D DCA map from the P2 detector reconstructed using counts of individual pixels mapped as a function of each scan step. A close look at the DCA map reveals pixel non-uniformities particularly at the edges and corners of most of these DCUs. These were first detected in 1D scans\cite{subpix} of the P2 detector plane but are now reproduced in much greater detail. The missing pixels in each of the DCUs are due to conductive epoxy bond separation between the CZT and the ASIC, likely due to the force of insertion of the elastomeric connectors (see figure \ref{intro2} and section 3 of Ref. \citenum{Hong2} for details on elastomeric connectors.). Figure \ref{pixel_map} (right) shows the response map zoomed in all the way to a single 600 $\mu m$ $\times$ 600 $\mu m$ pixel divided into 36 $\times$ 100 $\mu m$ $\times$ 100 $\mu m$ sub-pixels. Figure \ref{pixel_map} does not account for changes in Mini-X output flux as described in Section \ref{MiniX_perform} nor does it account for changes in flux rates in the detector due to a non-uniform hole pattern as discussed in Section \ref{initial_scan}. 

\begin{figure}[htp]
\centering
\includegraphics[width = 0.95\textwidth]{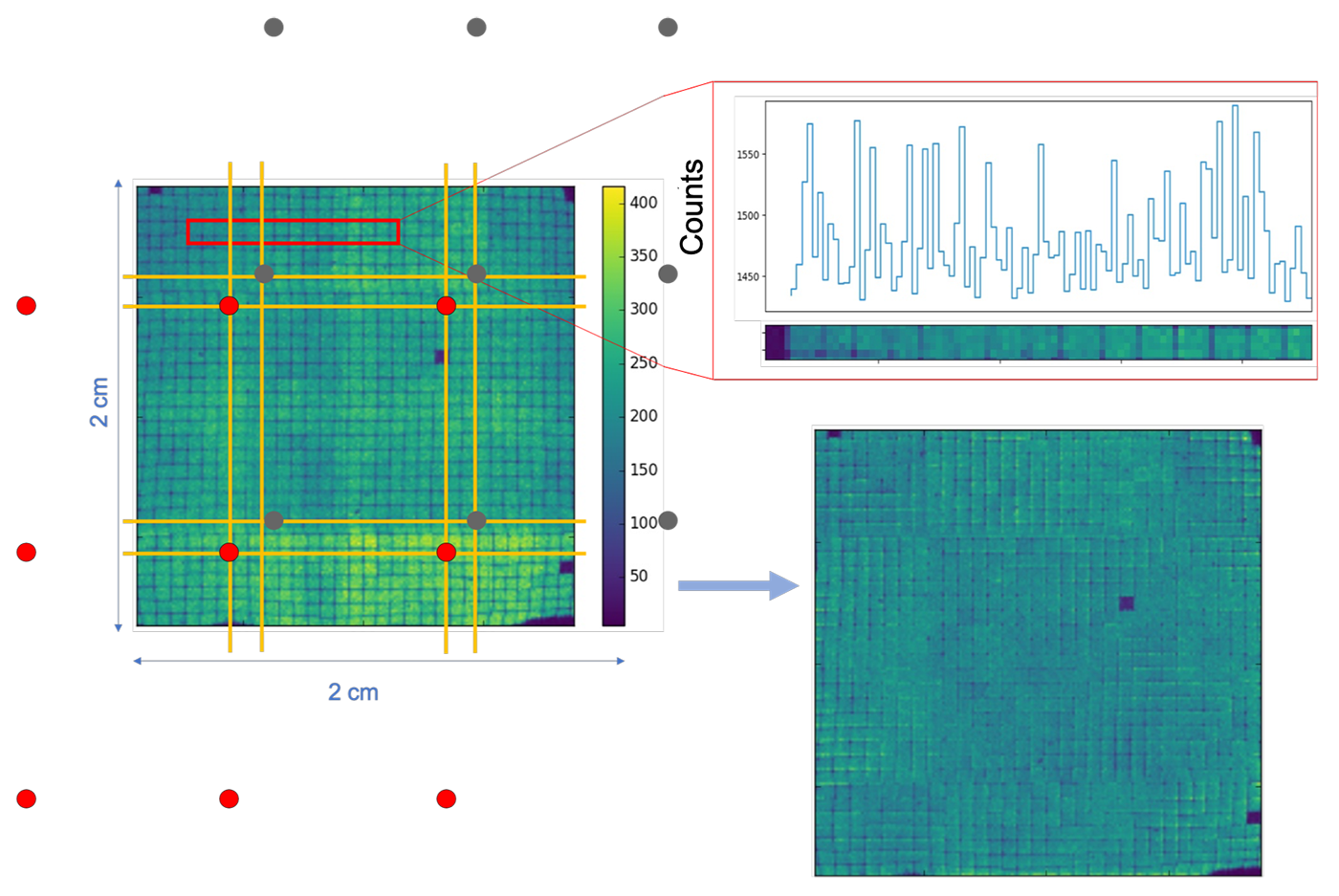}
\caption{Left: Raw $2cm \times 2cm$ DCU map. The yellow lines represent the overlapping area scanned by more than one hole. The red dots represent the initial mask pattern whereas the grey ones represent the mask pattern once the scan is complete. Top Right: Zoomed in image showing the variation in total flux as measured from the X-123 SDD.  Bottom Right: Clean DCU map after normalization of flux from the X-123 SDD data and flux from overlapping regions in the raw DCU map. }
\label{clean_DCU}
\end{figure}

Data from the X-123 SDD for each scan step was then used to normalize the output flux from Mini-X. Moreover, the movement of the XY-stage ensures that a particular portion of the DCU is scanned by a single hole with a $\simeq 2$ pixel overlap with an adjacent hole represented by the yellow lines in Figure \ref{clean_DCU}. The overlapping regions are used to normalize the flux from each hole thereby generating a much cleaner DCU map as shown in figure \ref{clean_DCU} (bottom right). Companion papers ~\citenum{basak1, subpix} analyzes the P2 detector response map in much greater detail - characterizing effective pixel size for each pixel\cite{basak1, subpix} in the P2 detector plane, determining pixel morphologies for each DCU or sections of DCUs to gauge pixel deformities and evaluate offsets for each pixel from its centroid\cite{basak1}.

The next generation HREXI detectors are slated to have much shorter scan times using HCF through changes in firmware and significant improvements in the front end data acquisition system. These include eliminating the ELBs altogether and replacing them with separate DCA boards and reading them out through a separate FPGA Mezzanine Board (FMB) among others\cite{josh1}.  This would enable us to use the  mask assembly in a configuration similar to figure \ref{mask1} (left) i.e. eight 100 $\mu m$ $\times$ 100 $\mu m$ holes over a single DCU instead of the current configuration of using four holes per DCU. Also, the use of TSV-ASIC in HREXI would enable flip-chip bonding between the back-end electronics and the ASIC thereby reducing the missing pixels which can be seen the P2 detector plane due to the conductive epoxy bond separation between the CZT and the ASIC\cite{Hong1}.

The presence of pixel non-uniformities, which were detected first in the 1D scan\cite{subpix} in the current 2D scan maps provide an independent verification of the scan completed using HCF. This along with the decrease in the scanning timescale from $\simeq$ 3 weeks to $\simeq$ 3 days and even shorter timescales for HREXI detectors clearly demonstrates the importance and utility of HCF for rapid evaluation and characterization of finely pixelated detectors at a sub-pixel level in a large imaging detector plane.   


\section{Acknowledgements}
This work was supported by NASA grant NNX17AE62G. The authors would like to thank Stan G. Corteau and Mike Mckenna of Harvard John A. Paulson School of Engineering and Applied Sciences machine shop for their support during the fabrication of HCF. 

\bibliography{report}   
\bibliographystyle{spiejour}   

\vspace{1ex}
\noindent Biographies and photographs of the authors are not available.

\listoffigures

\end{spacing}
\end{document}